\documentclass[twocolumn,prb]{revtex4}
\usepackage{epsfig,amssymb,amsmath}
\usepackage[english]{babel}

\begin{document}
\title{Peculiarities of phase dynamics of the Josephson junctions stack with the topologically nontrivial barriers}
\author{I. R. Rahmonov~$^{1,3}$}
\author{Yu. M. Shukrinov~$^{1,2}$}

\address{$^{1}$ BLTP, JINR, Dubna, Moscow Region, 141980, Russia \\
$^{2}$ Dubna State University, Dubna,  141980, Russia\\
$^{3}$Umarov Physical Technical Institute, TAS, Dushanbe, 734063 Tajikistan
}

\date{\today}

\begin{abstract}
The peculiarities of the phase dynamics of stacks of coupled Josephson junctions with topologically trivial and nontrivial barriers
have been investigated numerically and their comparative analysis is carried out. The effect of coupling and dissipation parameters on the parametric resonance in the breakpoint region is shown. It is found that the dependence of the breakpoint voltage on the dissipation parameter demonstrates a minimum due to the changing of frequency of longitudinal plasma wave. We have shown that in case of the stack with nontrivial barriers the observed minimum shifts along the $\beta$ to the $\sqrt{2}$ in comparison with trivial barriers stack. We assume that the found features may be used for the experimental determination of Majorana fermions in the stack of JJs with the nontrivial barriers.
\end{abstract}
\maketitle

\section*{Introduction}

The Majorana fermions~\cite{majorana37} are attracting considerable interest because of their importance for quantum computations. They are predicted to exist in Josephson junctions with topologically nontrivial barriers~\cite{fu08,tanaka09}. It is assumed that the topologically nontrivial states are formed on the boundary or surface of a topological insulator~\cite{fu07} and a semiconductor nanowire in the presence of the Rashba spin-orbit coupling and the Zeeman field~\cite{sau10}. The formation of Majorana states in Josephson junction leads to tunneling of quasiparticles with charge $e$ in compare with $2e$ in usual case~\cite{kitaev01, veldhorst12}. As a result~\cite{fu09}, the oscillation period of the Josephson current is doubled $I_{s}=I_{c}\sin{\varphi/2}$. This $4\pi$--periodicity was discussed by A. Kitaev in Ref.~\cite{kitaev01}, where an experimental observation of Majorana fermions was suggested by the investigation of quantum wire bridge between two superconductors. Later Kwon at. el.~\cite{kwon04} have demonstrated a fractional ac Josephson effect, which confirms the $4\pi$ periodicity. The tunneling conductance peak at zero voltage was observed experimentally for the first time in the superconductor -- semiconductor nanowire junction~\cite{mourik12}, which hosts the Majorana fermions.

Presence of the Majorana fermions in the systems containing the Josephson junctions has been
shown recently years~\cite{kitaev01,kwon04,fu09,maiti15,mourik12,wimmer10,potter10,veldhorst12,rahmonov-jetpl16,rahmonov-fnt17}. The majority of those works were concentrated on the studying of the single JJ hosted Majorana subgap states\cite{kitaev01,kwon04,maiti15} or SQUIDs with the nontrivial barriers\cite{veldhorst12,rahmonov-jetpl16,rahmonov-fnt17}. The possibility of appearance of subharmonic odd Shapiro steps is shown in Ref.\cite{maiti15}. In addition, it is shown that the JJ hosting Majorana bound states also display an additional sequence of steps in the devil's staircase structure seen in their I-V characteristics. In Ref.~\cite{fu09} has been shown that the DC-SQUIDs with the nontrivial barriers are expected to be used as a real quantum gates, since Majorana fermions exhibit a non-Abelian statistics which leads to a topological protection to errors. An optimization study for Majorana fermions in a DC-SQUID with topologically nontrivial barriers was performed by Veldhorst et al.~\cite{veldhorst12}. In our recent work~\cite{rahmonov-jetpl16} we have shown the effect of $4\pi$ periodic Josephson current on the return current and resonance features of the DC--SQUID.

It was demonstrated that the fractional Josephson effect in the junctions with the nontrivial barriers is one of the perspective methods of Majorana fermion detection\cite{houzet13, dominguez12}.

However, there is another interesting opportunity for the manifestation of the Majorana fermions in Josephson junctions systems. Investigation of a stack of coupled JJs with the Majorana edge states is of great interest, since it could provide a new method to detect those states. The stack of coupled Josephson junctions has rich nonlinear properties. An actual problems is to investigate such stacks with the nontrivial barriers, which can host Majorana edge states. As it is well known in the IV--curve of the stack of Josephson junctions a breakpoint due to the creation of longitudinal plasma wave and realization of the parametric resonance is observed. It is interesting to clarify the breakpoint region features in IV-curve of the stack of JJs with nontrivial barriers. In this work we present the results of detailed numerical investigations of the phase dynamics of the stacks of Josephson junctions with the trivial and nontrivial barriers. The comparative analysis of stacks with the topologically trivial and nontrivial barriers are carried out.

\section{Theoretical model and formulation}

Let us consider a stack of Josephson junctions with topologically nontrivial barriers. As was mentioned in the introduction, the presence of Majorana fermions leads to the single electron tunneling and doubles of the period of phase difference of the order parameter\cite{kitaev01,veldhorst12,fu09}. Therefore, within a CCJJ+DC model\cite{sm-prl07,sms-prb08,rahmonov-jetp12} (taking into account the existence of Majorana fermions) it is sufficient to replace $2e$ by $e$ and $\varphi$ by $\varphi/2$ in the corresponding terms of the equations. Thus, for both trivial and nontrivial cases, the Josephson relation is the same
\begin{equation}
\label{Jos_rel}
\frac{\hbar}{e}\frac{d( \varphi_{l}/2)}{dt}=\frac{\hbar}{2e}\frac{d \varphi_{l}}{dt}=V_{l}-\alpha(V_{l+1}+V_{l-1}-2V_{l})
\end{equation}

\noindent where $\varphi_{l}$ and $V_{l}$ are the phase difference and voltage across $l$th JJ, respectively, $\alpha$ is the capacitive coupling parameter. The current $I$ passing trough each JJ of the stack can be written as the following

\begin{equation}
\label{system_eq1}
\displaystyle I=C\frac{\partial V_{l}}{\partial t}+\frac{\hbar}{2eR}\frac{\partial \varphi_{l}}{\partial t}+ I_{c}\big[\varepsilon\sin\varphi_{l} +(1-\varepsilon)\sin\frac{\varphi_{l}}{2}\big]
\end{equation}

\noindent where $C$ is a capacitance, $R$ is a resistance and $I_{c}$ is a critical current of JJ. We note that to obtain the general system of equations for both type of stacks with trivial and nontrivial barriers, we have used the expression $I_{s}=I_{c}\big[\varepsilon\sin\varphi_{l} +(1-\varepsilon)\sin\frac{\varphi_{l}}{2}\big]$ for superconducting current, where $\varepsilon$ is the ratio of $2\pi$ and $4\pi$ periodic Josephson currents through the junctions, i.e. $\varepsilon=1$ and $\varepsilon=0$ relate to the trivial and nontrivial cases, respectively. The ratio parameter $\varepsilon$ was first introduced by Veldhorst with coauthors~\cite{veldhorst12}.

Using Josephson relation (\ref{Jos_rel}) and expression for current (\ref{system_eq1}), we write the system of equations in normalized units as following, 

\begin{equation}
\label{eq_sys0}
\left\{\begin{array}{ll}
\displaystyle \frac{d\varphi_{l}}{d t}=V_{l}-\alpha(V_{l+1}+V_{l-1}-2V_{l})
\vspace{0.2 cm}\\
\displaystyle \frac{d V_{l}}{dt}=I-\varepsilon\sin(\varphi_{l})-(1-\varepsilon)\sin(\frac{\varphi_{l}}{2}) -\beta \frac{d\varphi_{l}}{dt}
\end{array}\right.
\end{equation}

which describes the dynamics of the stack of JJs, where $\beta =1/R\sqrt{\hbar/(2eI_{c}C)}$ is a dissipation parameter. In the system of equations (\ref{eq_sys0}) time is normalized to $\omega_{p}^{-1}$, where $\omega_{p}=\sqrt{2eI_{c}/(\hbar C)}$, voltage -- to the $V_{0}=\hbar \omega_{p}/2e$ and bias current $I$ -- to the critical current $I_{c}$.

In our calculations the bias current is increased from $I_{0}=0.1$ till $I_{max}$ and further reduced to zero by step of $\Delta I=0.0005$.
At each fixed value of the bias current, the system of differential equations (\ref{eq_sys0}) are solved by the fourth order Runge-Kutta method in time interval from $0$ to $T_{max}=800$ with a step $\Delta t=0.05$. As a result, we have obtained a voltage $V_{l}$ and a phase difference $\varphi_{l}$ as a functions of time. Then, the obtained values of the voltage $V_{l}$ is averaged in a time interval [50,800].
The electric charge in the superconducting layer is calculated using the expression $Q=Q_{0}(V_{l+1}-V_{l})$\cite{rahmonov-jetp12, shukrinov-lncs12}, where $Q_{0}=\varepsilon\varepsilon_{0}V_{0}/r^{2}_{D}$. The details of simulation method is presented in the previous works\cite{rahmonov-jetp12, shukrinov-lncs12}.

\section{Results and discussion}

\begin{figure}[h!]
 \centering
 \includegraphics[height=50mm]{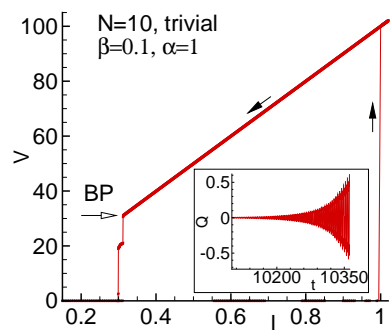}
\caption{Current--voltage characteristics of the stack of 10 coupled JJs with the topologically trivial barriers, simulated for $\beta=0.1$ and $\alpha=1$. The inset shows the part of the time dependence of the charge in the superconducting layer at the breakpoint region, which demonstrates the increasing of the amplitude of the charge oscillation.}
\label{cvc_time_dep}
\end{figure}

Let us first of all discuss the main features of the IV--characteristics of stack with topologically trivial barriers. The simulated IV-curve is shown in Fig.\ref{cvc_time_dep}. The calculation of IV--curve is performed by the increasing of bias current from the zero to $I_{max}=2$ and then decreased to the zero. The arrows show the direction of bias current changing. The IV--curve demonstrates a hysteresis and a breakpoint region in the outermost branch\cite{sm-prl07,smp-prb07,sms-prb08}. The position of the breakpoint is shown by the horizontal hollow arrow and it is marked as BP. The breakpoint phenomena in the stack of JJs with the trivial barriers was discussed in the several works\cite{sm-prl07,smp-prb07,sms-prb08,iso-apl08}. It was shown that in this region the longitudinal plasma wave along the stack creates and due to the parametric resonance, the amplitude of charge oscillations in the superconducting layers increases exponentially. In the inset of the Fig.\ref{cvc_time_dep} we show the time dependence of the charge in the superconducting layer, which demonstrates an increasing of the charge amplitude due to the parametric resonance.

The character of the charge oscillation corresponding to the bias current value $I=0.312$ and voltage $V=30.6$ is shown in Fig.\ref{time_dep_fft}(a).  In this case, the Josephson frequency is equal to $\omega_{J}=3.06$ ($f_{J}=0.487$). As we can see, the charge oscillations show the non-harmonic behavior. The results of the FFT analysis of this time dependence is shown in Fig.\ref{time_dep_fft}(b), which demonstrates the frequency of the formed longitudinal plasma wave $f_{LPW}=0.245$. It is two time less than Josephson frequency $f_{J}$. The additional pick approximately corresponds to the sum of the LPW and Josephson frequency.

\begin{figure}[h!]
 \centering
 \includegraphics[height=30mm]{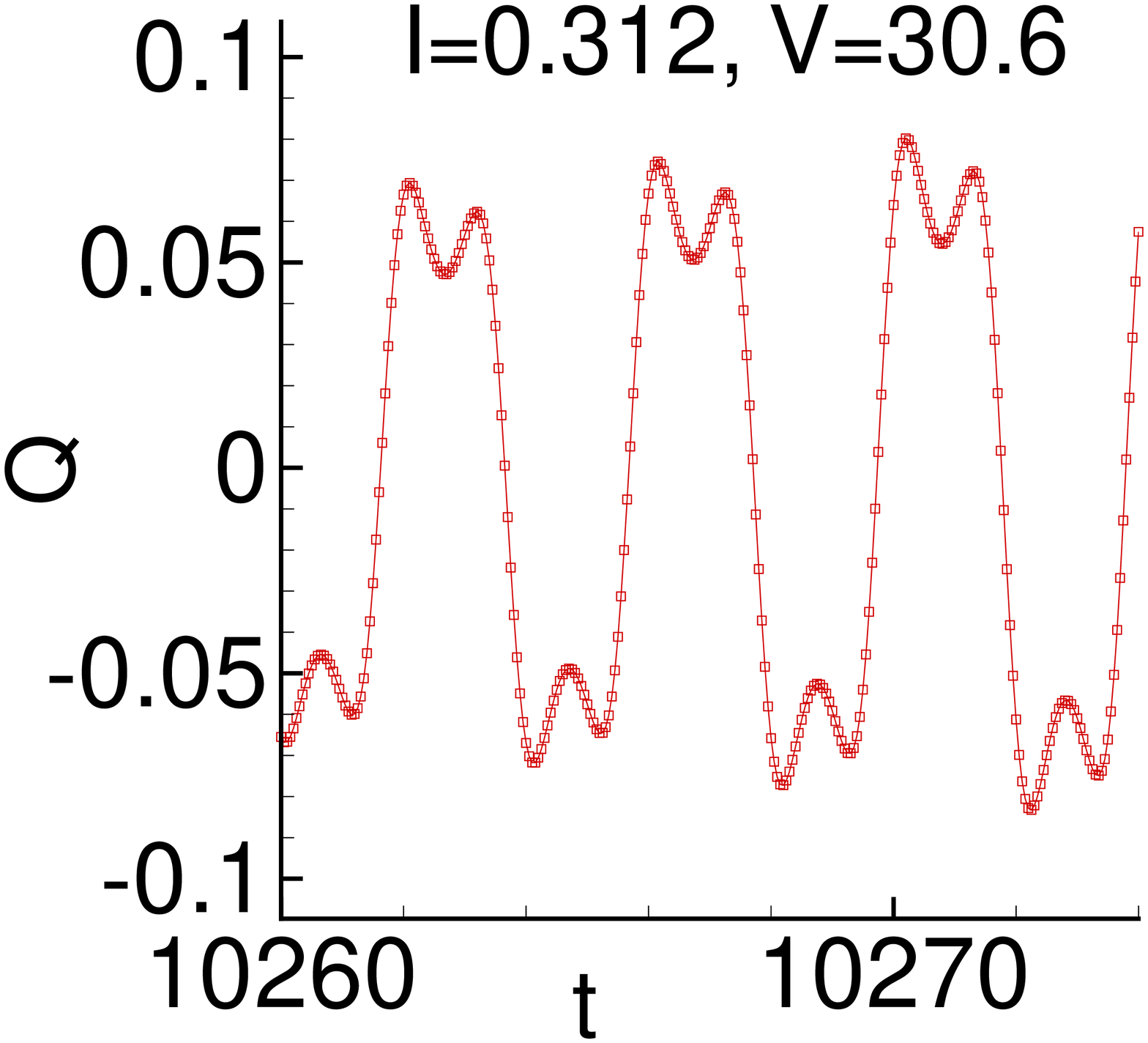}\hspace{0.3cm}
  \includegraphics[height=30mm]{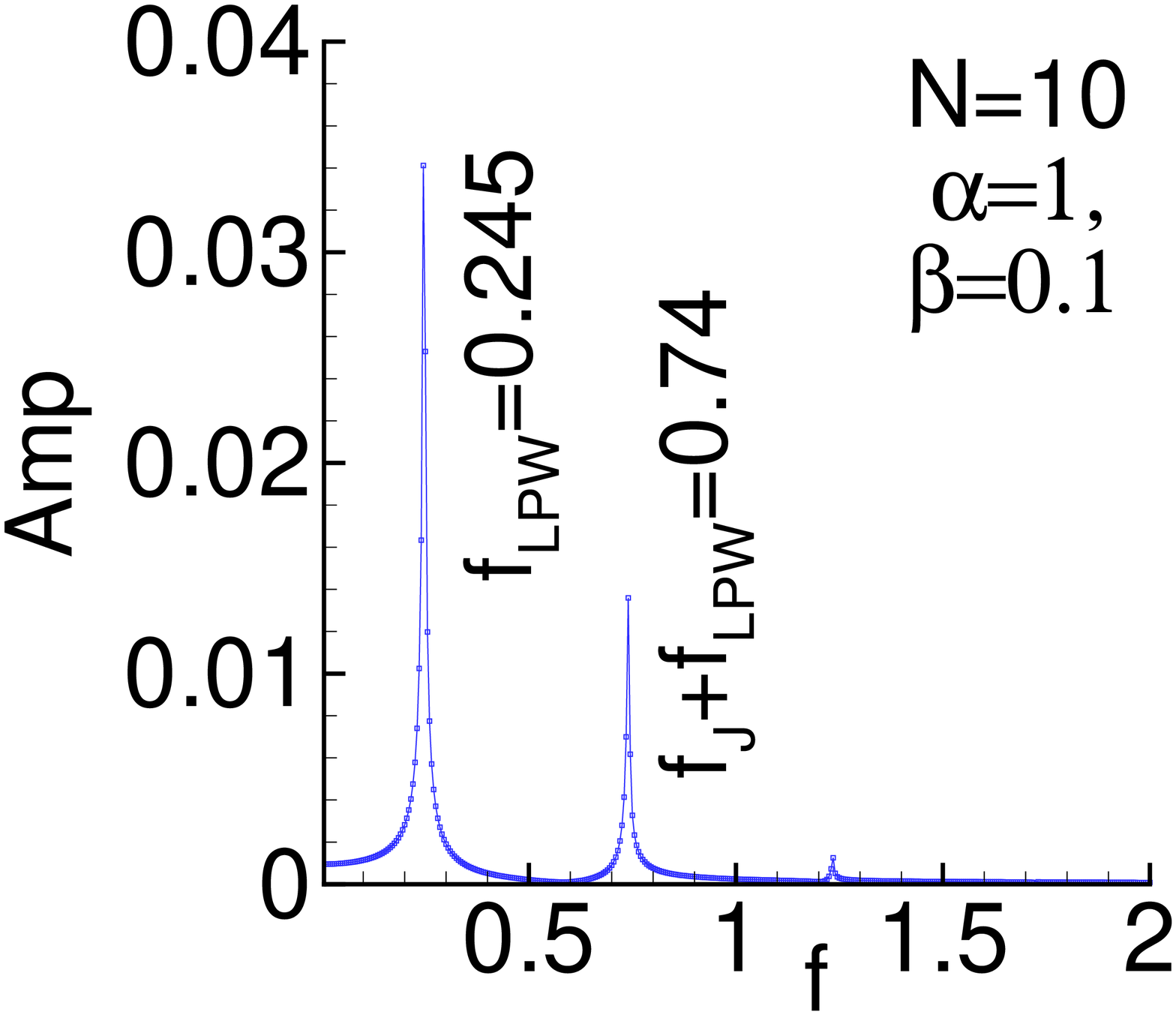}
\caption{(a) The time dependence of the electric charge in the superconducting layer corresponding to the breakpoint region; (b) The FFT analysis for the charge time dependence shown in (a).}
\label{time_dep_fft}
\end{figure}

\begin{figure}[h!]
 \centering
 \includegraphics[height=50mm]{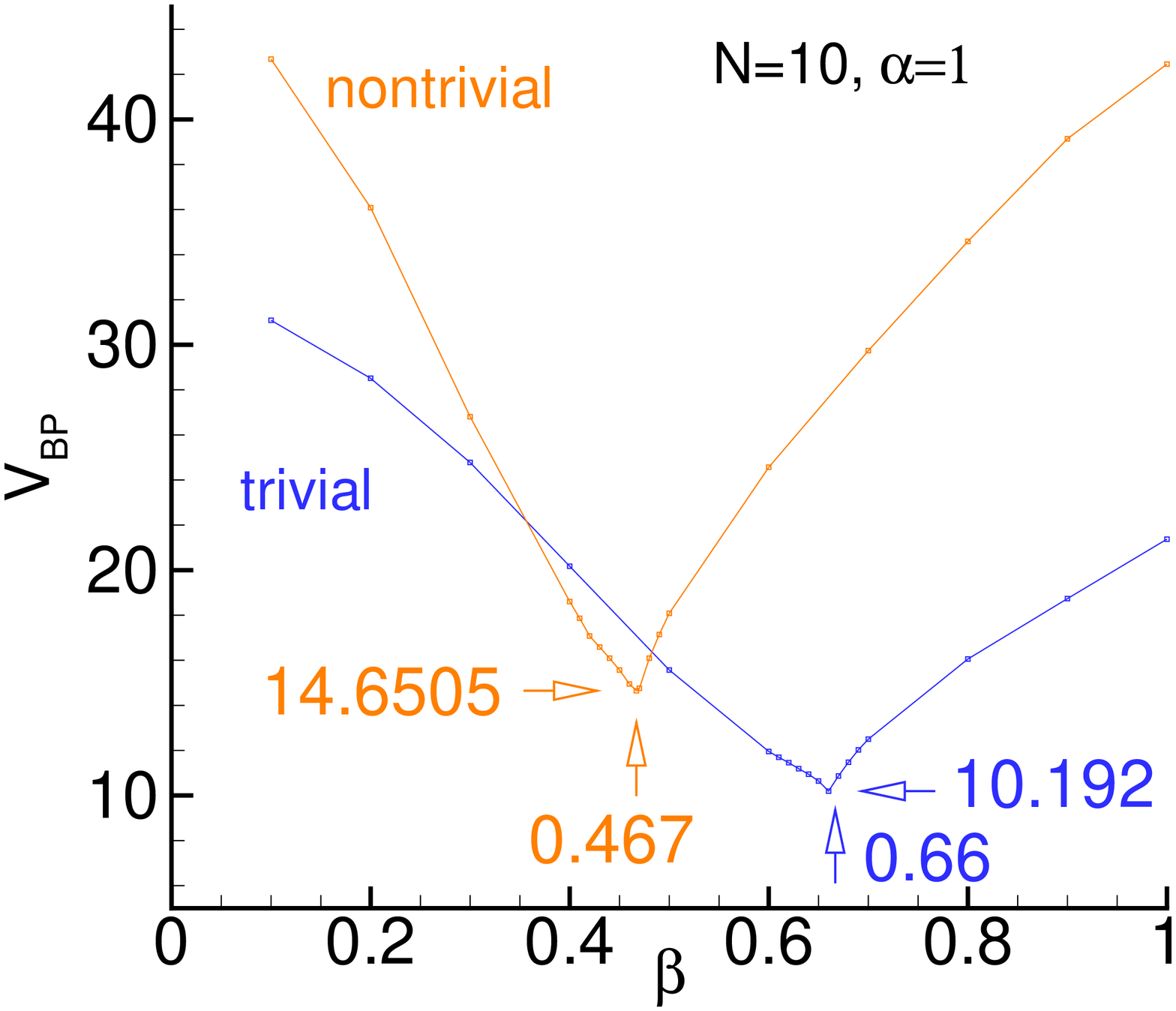}
\caption{$V_{BP}$ depending on the $\beta$ for the stack with the trivial and nontrivial barriers.}
\label{V_bp}
\end{figure}

In case of the stack with the nontrivial barriers we obtain the qualitatively similar behavior. Actually, IV--curve of the JJs stack with nontrivial barriers also demonstrates the breakpoint and parametric resonance. But at the same chosen parameters, the position of breakpoint and frequency, corresponded to the parametric resonance, are different. In Fig.\ref{V_bp} we show the dependencies of the voltages $V_{BP}$ corresponded to the breakpoint on the $\beta$, in both trivial and nontrivial cases. It can be seen that these both dependencies demonstrate the minimum. For the JJs stack with trivial barriers, minimum of breakpoint voltage is equal to $V_{BP}=10.192$ and it is observed for $\beta=0.66$. For the stack with the nontrivial barriers the corresponding minimum is observed at $V_{BP}=14.6505$ and $\beta=0.467$.  We stress that in nontrivial case the above mentioned minimum is observed for the $\sqrt{2}$ less beta value, in compare with the trivial case. Also, the breakpoint voltage $V_{BP}$, corresponding to the minimum for the stack with nontrivial barriers is $\sqrt{2}$ time larger than the $V_{BP}$ in the stack with trivial barriers.

Let us discuss now the origin of the observed minimum in the $\beta$--dependence of $V_{BP}$. We will do it just for the trivial case. In Fig.\ref{time_dep_fft}(a) we show the time dependence of the charge in the superconducting layer before the above mentioned minimum i.e. for the value of the dissipation parameter $\beta=0.65$. The results of FFT analysis of this time dependence, which is presented in Fig.\ref{time_dep_fft}(b), demonstrates two frequencies: one corresponds to the half of Josephson frequency and has big amplitude, and second one corresponds to the Josephson frequency with a small amplitudes. For the value of $\beta$ larger than the minimum value, the behavior of charge oscillation changes in compare with a case before minimum. It can be seen in the time dependence of charge for the $\beta=0.67$, which is shown in Fig.\ref{time_dep_fft}(c).  Figure \ref{time_dep_fft}(d) demonstrates the results of FFT analysis of charge--time dependence for $\beta=0.67$. Here we can see, that the frequency of LPW coincides with the Josephson frequency, and pick corresponded to the half Josephson frequency disappears. So that is the reason of manifestation of observed minimum.
\begin{figure}[h!]
 \centering
 \includegraphics[height=30mm]{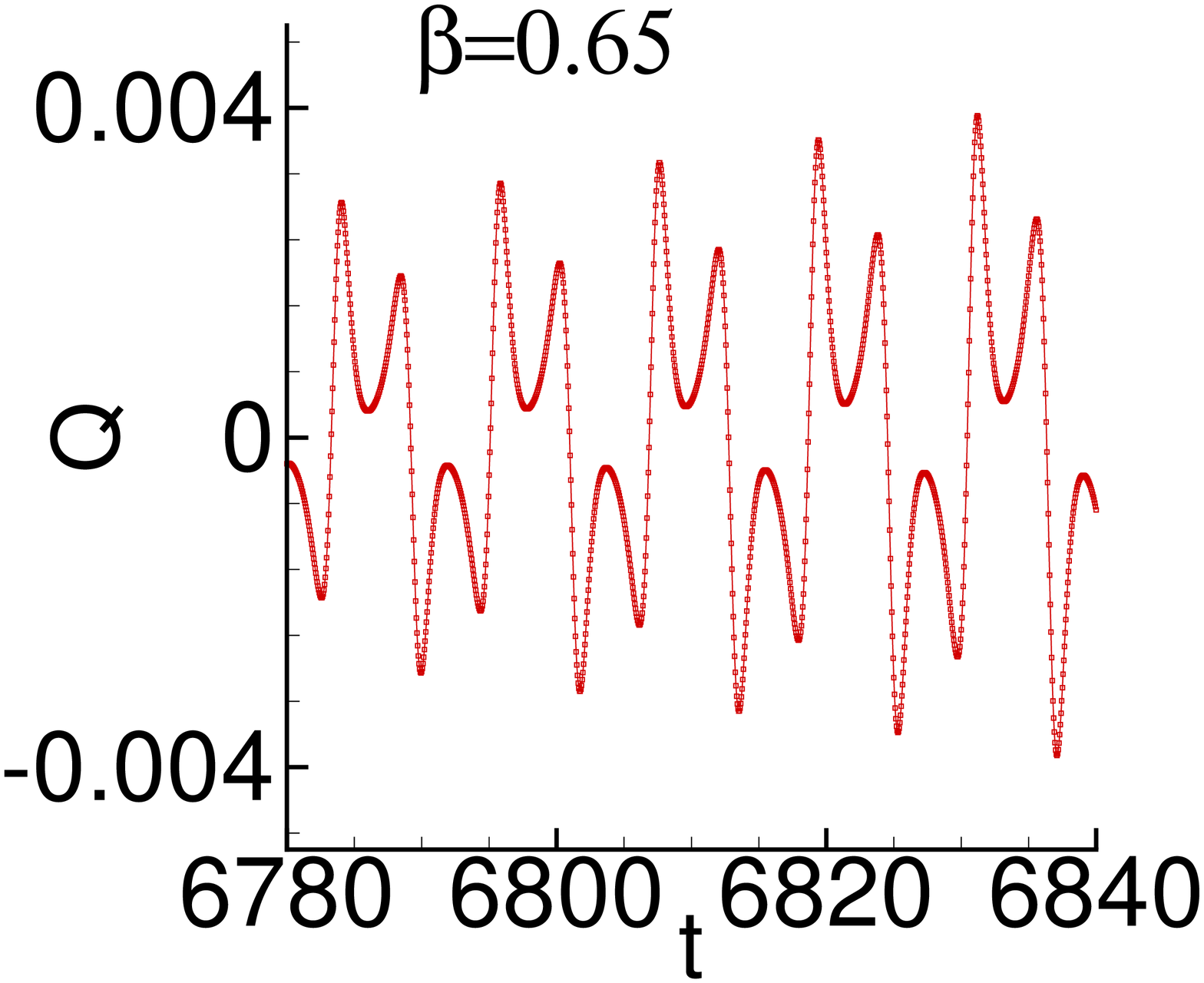}\hspace{0.3cm}
 \includegraphics[height=30mm]{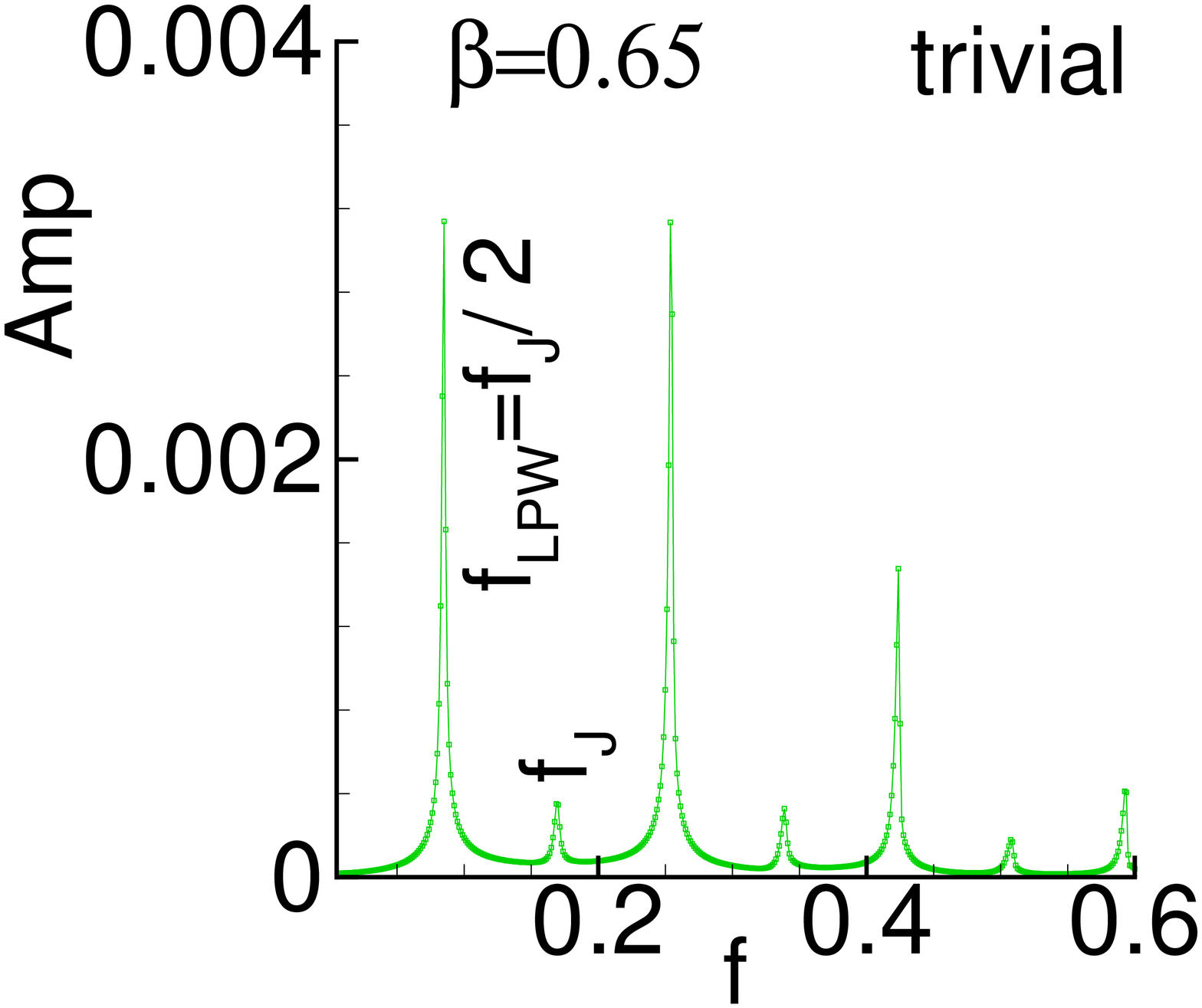}\\
  \includegraphics[height=30mm]{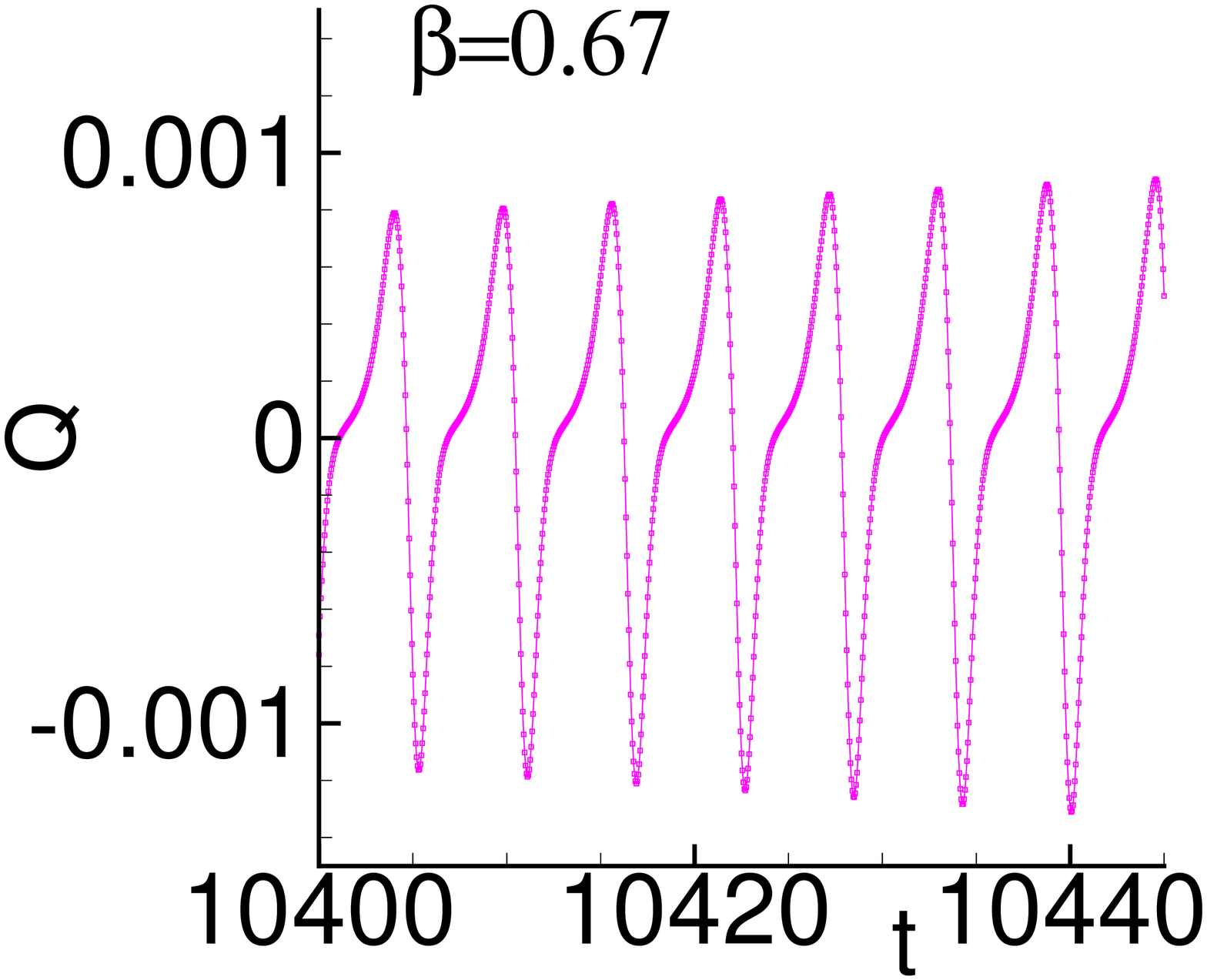}\hspace{0.3cm}
 \includegraphics[height=30mm]{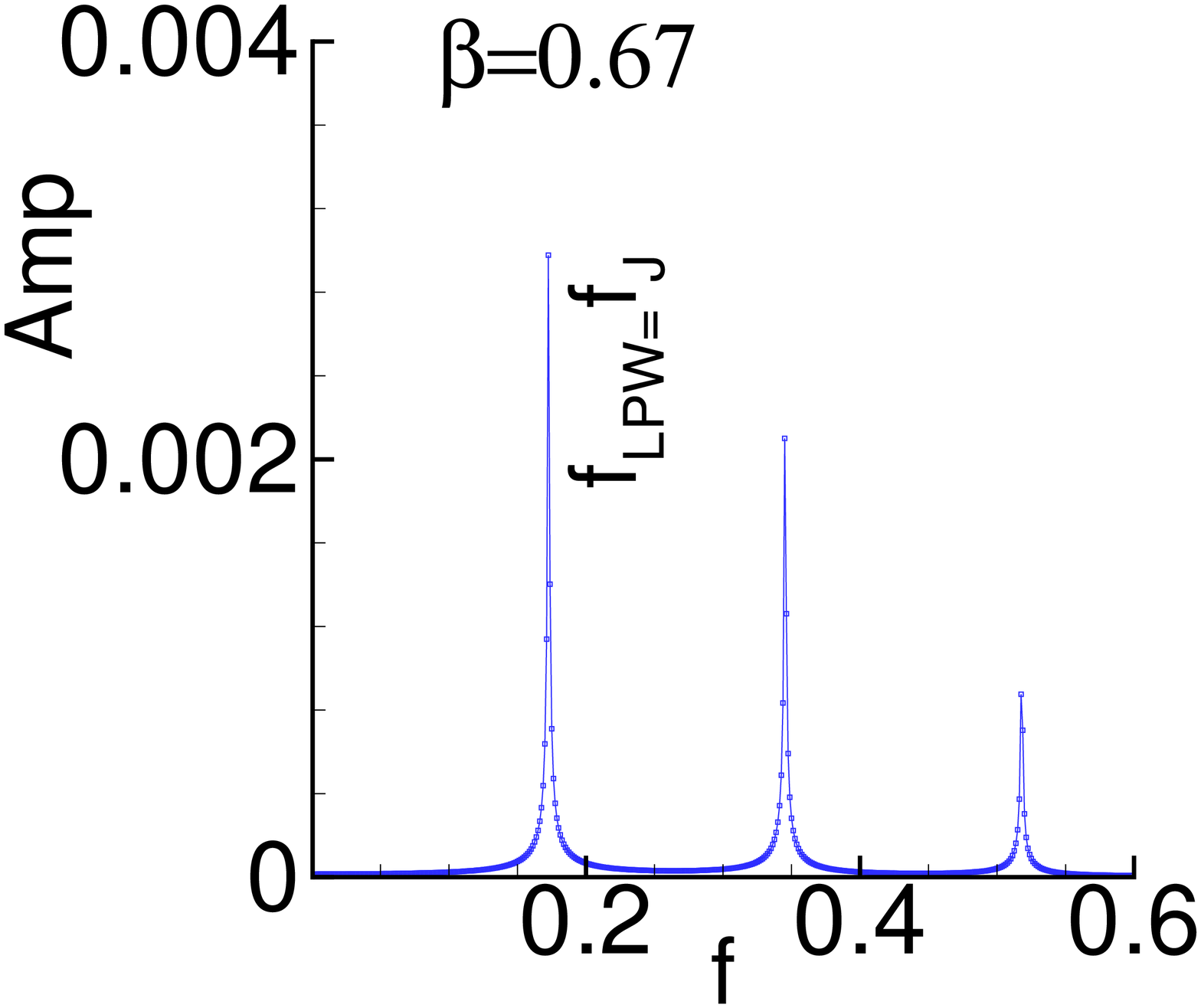}\\
\caption{(a) Time dependence of electric charge at $\beta=0.65$; (b) The FFT analysis of the time dependence shown in (a); (c) The same as (a) at $\beta=0.67$; (d) The FFT analysis of the time dependence (c).}
\label{time_dep_fft}
\end{figure}

\section{Conclusion}
The peculiarities of the phase dynamics of stack of coupled Josephson junctions with trivial and nontrivial barriers
have been studied numerically. Results of the detailed analysis of effect of  dissipation parameter on the breakpoint region in the IV--characteristic of JJs stack are presented. We have shown that the dissipation parameter lead to minimum in $\beta$--dependence of the breakpoint voltage $V_{BP}$. The performed analysis shows that the observed minimum corresponds to the case when oscillations with the half Josephson frequency in the charge--time dependence are disappeared. The main results is that in case of the stack with nontrivial barrier the observed minimum shifts along the $\beta$ to the value $\sqrt{2}$. We consider that this fact may be used for the experimental determination of majorana fermions in the stack of JJs.

\section{Acknowledgements}
The reported study was funded by RFBR according to the research project 16-52-45011 and by the grant of AYSS of JINR with the project 18-302-08.

{}

\end{document}